\documentclass[aps,prb,showpacs,preprintnumbers,twocolumn,amsmath,superscriptaddress]{revtex4}

\usepackage{graphicx}
\usepackage{color}

\begin{document}
\topmargin 0.1in

\newcommand{\ket}    [1]{{|#1\rangle}}
\newcommand{\bra}    [1]{{\langle#1|}}
\newcommand{\braket} [2]{{\langle#1|#2\rangle}}
\newcommand{\bracket}[3]{{\langle#1|#2|#3\rangle}}

\def\bea{\begin{eqnarray}}
\def\nn{\nonumber\\}
\def\eea{\end{eqnarray}}
\def\beq{\begin{equation}}
\def\eeq{\end{equation}}

\def\h0{H_k^0}
\def\ee{\mathcal E}
\def\vv{v'}
\def\dd{\widetilde{\partial}_k}

\title{Nonadiabatic wavepacket dynamics: $k$-space formulation}

\author{J. M. Pruneda}
\affiliation{Department of Physics, University of California,
Berkeley, CA 94720, USA}
\affiliation{Institut de Ciencia de Materials de Barcelona,
CSIC Campus U.A.B., 08193 Bellaterra, Barcelona, Spain}
\author{Ivo Souza}
\affiliation{Department of Physics, University of California,
Berkeley, CA 94720, USA}

\date{\today}

\begin{abstract}
The time evolution of wavepackets in crystals in the presence of a homogeneous 
electric field is formulated in $k$-space 
in a numerically tractable form. 
The dynamics is governed by separate equations for the 
motion of the waveform in $k$-space and for the evolution 
of the underlying Bloch-like states. 
A one-dimensional
tight-binding model is studied numerically, and both Bloch oscillations and 
Zener tunneling are observed.
The long-lived Bloch oscillations of the wavepacket center under weak fields 
are accompanied by oscillations in its spatial spread.
These are analyzed in terms of a $k$-space expression for the spread having
contributions from both the quantum metric and the Berry
connection of the Bloch states. 
We find that when sizeable spread oscillations do occur, they are mostly due to
the latter term.
\end{abstract}


\maketitle

\section{Introduction}

The study of the dynamics of electron wavepackets in crystals 
has experienced a revival in recent 
years.  The development of heterostructure superlattices, photonic crystals, 
and optical lattices
has opened new possibilities for the experimental realization of 
fundamental dynamical
effects such as Bloch 
oscillations\cite{superlattices,Dahan,opticalBO,opticalBO2,BO-BEC} and 
Zener tunneling.\cite{Niu96,opticalZener,Zener-BEC}
The wavepacket picture of transport has also shed light
on subtle transport phenomena in solids. 
For instance, the intrinsic anomalous Hall 
effect in ferromagnets was shown to result from a Berry-curvature 
term in the wavepacket group velocity.\cite{chang08}

Numerical simulations provide valuable insights into the dynamics of 
wavepackets in crystals. For instance,
Bouchard and Luban\cite{BouchardLuban} carried out a detailed study
on a one-dimensional biased lattice,
finding a rich variety of dynamical phenomena
(Bloch oscillations of the center of mass, coherent breathing modes, 
Zener tunneling, and intrawell oscillations) as a function of the field-free 
band structure, field strength,
and the form of the initial wavepacket. In order to
solve numerically the time-dependent Schr\"odinger equation, 
they employed a supercell geometry with hard-wall boundary conditions;
care had to be taken to ensure that the wavepacket never
came close to the hard-wall boundaries for the duration of the simulation.
In situations where unbounded acceleration (via Zener tunneling)
of a significant portion of the 
wavepacket takes place, a large supercell must then be used, which
may become computationally demanding. In principle that can be avoided by
switching from hard-wall to periodic boundary conditions.
However, the inclusion in the Hamiltonian of the nonperiodic electric
field term $e\boldsymbol{\ee}\cdot{\bf r}$ then
becomes problematic. 
A successful numerical strategy 
for describing homogeneous electric fields under periodic boundary conditions  
was developed in
Refs.~\onlinecite{SIV02,umari02} for static fields, 
and generalized to time-dependent fields in Ref.~\onlinecite{SIV}.

In Refs.~\onlinecite{SIV02,umari02,SIV} the goal was to
solve for the electronic structure of insulators in the presence of a 
homogeneous field. 
In this work we use a similar strategy to describe wavepacket dynamics.
Our starting point is to 
express the wavepacket as a linear superposition of Bloch states,
\beq
\label{wavepacket}
\ket{\phi}= \int_0^{2\pi/a} dk\, f_k \ket{\psi_k}=
\int_0^{2\pi/a} dk\, e^{ikx} f_k\ket{v_k}
\eeq
(for simplicity we shall work in one dimension), and to follow the
time evolution of both the waveform $f_k$ and the underlying states 
$\ket{v_k}$.
If a wavepacket, initially prepared in a given band, is constrained to 
remain in the same band at later times,
one obtains the ``semiclassical'' approximation,
which becomes exact in the adiabatic limit.
Instead, we will allow for a fully unconstrained time evolution.
As a result, for $t>0$ the states $\ket{\psi_k}$ may become an 
admixture $\sum_n c_{nk}\ket{\psi_{nk}^{(0)}}$
of several eigenstates of the crystal Hamiltonian $H^0$. 
We shall refer to such nonadiabatic
states as Bloch-like, since they retain the Bloch form, with $v_k(x+a)=v_k(x)$.

The manuscript is organized as follows. 
The equations of motion for $f_k$ and $\ket{v_k}$, and the expressions 
for the center and spread of the packet in terms of them, are derived in 
Secs.~\ref{sec:method} and \ref{sec:center_spread} respectively.
In Sec.~\ref{sec:results} we perform simulations for a one-dimensional 
tight-binding Hamiltonian, using a numerically tractable form, 
given in the Appendix, of those equations.

\section{Dynamical equations in $k$-space}
\label{sec:method} 

The wavepacket evolves 
according to the Schr\"odinger equation ($e=\hbar=1$)
\beq
\label{eq:tdse}
i\frac{d\ket{\phi}}{dt}=(H^0+\ee x)\ket{\phi},
\eeq
where 
$\ee$ is the electric field, which can be time-dependent but must be
spatially uniform. 
Since $f_k$ and $\ket{v_k}$ 
enter Eq.~(\ref{wavepacket}) as a product,
they are individually defined only up to a multiplicative factor which, because
$\bra{v_k}v_k\rangle=1$, must take the form $e^{i\varphi_k}$.
We fix this phase arbitrariness by choosing $f_k$ to be real and 
positive.

Inserting Eq.~(\ref{wavepacket}) into Eq.~(\ref{eq:tdse}),
\bea
\label{eq:manip}
&&i\int dk\,e^{ikx}
\big(
  f_k\ket{\dot{v}_k}+\dot{f}_k\ket{v_k}
\big)=\nn
&=&\int dk\,e^{ikx}
\h0 f_k \ket{v_k}
+\ee\int dk\,e^{ikx} x f_k \ket{v_k},
\eea
where $\h0=e^{-ikx}H^0 e^{ikx}$ and henceforth the integration range from 0 to 
$2\pi/a$ will be implied. Rewriting
\bea
&&\ee\int dk\left( -i\partial_k e^{ikx}\right)f_k \ket{v_k} \nn
&=&\ee\int dk\,e^{ikx}i\left(\ket{v_k}\partial_k f_k+
f_k\ket{\partial_kv_k}\right)
\eea
($\partial_k\equiv \partial/\partial k$ and
an integration by parts was performed) then yields, at each $k$, 
\begin{eqnarray}
\label{global}
if_k|\dot{v}_k\rangle + i\dot{f}_k|v_k\rangle &=& H_{k}^0f_k|v_k\rangle
+ \nn
&+& i{\mathcal E}\left(|v_k\rangle\partial_k f_k +f_k|\partial_kv_k\rangle
\right).
\end{eqnarray}
Contracting with $\bra{v_k}$ and subtracting from the 
resulting equation its complex conjugate, we arrive at the 
equation of motion for $f_k$,
\begin{equation}
\label{formevolution}
\dot{f}_k = {\mathcal E}\partial_k f_k,
\end{equation}
where the reality of $f_k$ was used, together with the relations 
$\langle\dot v_k|v_k\rangle=-\langle v_k|\dot v_k\rangle$ and
$\langle \partial_k v_k|v_k\rangle=-\langle v_k|\partial_k v_k\rangle$.
To find the equation of motion for $\ket{v_k}$ we
plug (\ref{formevolution}) back into (\ref{global}):
\begin{equation}
\label{stateevolution}
i\ket{\dot{v}_k}=\left(\h0+i\ee\partial_k\right)\ket{v_k}.
\end{equation}

Eqs.~(\ref{formevolution})--(\ref{stateevolution}) govern the
coherent wavepacket dynamics.
Eq.~(\ref{stateevolution}) was previously obtained in Ref.~\onlinecite{SIV}, 
where it was shown to describe the dynamics of valence electrons in
insulators under the homogeneous field $\ee(t)$. Here it describes the
nonadiabatic evolution of the Bloch-like states $\ket{v_k}$ supporting the 
wavepacket.
As for Eq.~(\ref{formevolution}), it
is the familiar result for the $k$-space dynamics of the waveform, 
which remains valid in the presence of interband mixing. 

For numerical implementation
the $k$-derivatives must be replaced by 
finite-difference expressions over a $k$-point mesh. 
While such discretization is straightforward
for Eq.~(\ref{formevolution}), Eq.~(\ref{stateevolution}) 
requires some care. As in Ref.~\onlinecite{SIV}, we replace it with
\begin{equation}
\label{state2evolution}
i\ket{\dot{\vv}_k}=
\big(\h0+ i\ee\widetilde{\partial}_k\big)
\ket{\vv_k},
\end{equation}
where $\ket{\dd \vv_k}\equiv Q_k\ket{\partial_k \vv_k}$ 
($Q_k=1-\ket{v_k}\bra{v_k}=1-P_k$).
Unlike $\ket{\partial_k v_k}$, $\ket{\dd \vv_k}$
lends itself to a numerically robust finite-differences representation
(see Appendix).

The states $\ket{\vv_k}$ obeying Eq.~(\ref{state2evolution}) differ
from the states $\ket{v_k}$ in Eq.~(\ref{stateevolution}) by a phase factor,
\beq
\label{eq:v_two}
\ket{\vv_k}=e^{i\beta_k}\ket{v_k}=U_k\ket{v_k},
\eeq
which we must keep track of.
Inserting (\ref{eq:v_two}) into (\ref{state2evolution}) and using 
(\ref{stateevolution}) yields
\begin{equation}
\label{phaseevolution}
i\dot{U}_k= -{\mathcal E}A_kU_k,
\end{equation}
where $A_k$ is the Berry connetion,
\beq
\label{eq:connection}
A_k=i\langle v_k|\partial_kv_k\rangle.
\eeq

Eqs.~(\ref{formevolution}), (\ref{state2evolution}) and (\ref{phaseevolution}) 
are the desired dynamical equations for $f_k$, $\ket{v_k'}$, and $U_k$. 
Together they determine the time evolution of the wavepacket
\beq
\ket{\phi}=\int dk\,f_ke^{ikx}U_k^*\ket{\vv_k}.
\eeq
In practice
Eqs.~(\ref{state2evolution}) and (\ref{eq:connection}) are replaced by 
the discretized forms
(\ref{eq:tdse_b}) and (\ref{eq:A_disc}), respectively.

\section{Wavepacket center and spread}
\label{sec:center_spread}

In the previous Section we formulated the dynamics of the wavepacket 
(\ref{wavepacket}) in terms of $f_k$ and $\ket{v_k}$.
Here we shall express its center and
spread in terms of those same $k$-space quantities.

\subsection{$k$-space expressions}

Let us define the generating function for the spatial 
distribution of the wavepacket:
\beq
\label{eq:gen}
C(q)=\langle \phi|e^{-iqx}|\phi\rangle=
\frac{2\pi}{a}\int dk\, f_k f_{k+q}\langle v_k|v_{k+q}\rangle,
\eeq
where the second equality follows from Eq.~(\ref{wavepacket}) together with the
identity
\beq
\label{eq:orthonorm}
\left<\psi_{k_1}\left|e^{-iqx}\right|\psi_{k_2}\right>=
\frac{2\pi}{a}\delta(k_2-k_1-q)
\langle v_{k_1}|v_{k_1+q}\rangle.
\eeq
The first moment is given by
\beq
\label{eq:x_avg}
\langle x\rangle =i\partial_q C(q)|_{q=0} 
=\langle iD_k\rangle=\langle A_k\rangle,
\eeq
where $iD_k$ is the Hermitian operator
\beq
iD_k=i\partial_k+A_k
\eeq
and we have introduced the notation
\beq 
\langle {\cal O}_k\rangle\equiv\frac{2\pi}{a}\int dk f_k {\cal O}_k f_k.
\eeq
The last equality in Eq.~(\ref{eq:x_avg}) follows from $f_k$ being real. 

Next we evaluate the spread 
\beq
(\Delta x)^2=\langle x^2\rangle-\langle x\rangle^2.
\eeq
For $\langle x\rangle^2$ we use (\ref{eq:x_avg}), while
$\langle x^2\rangle$ is given by $i^2\partial^2_q C(q)|_{q=0}$:
\beq
\langle x^2\rangle=
\frac{2\pi}{a}
\left[
  \int dk\,(\partial_k f_k)^2
 +\int dk\,f_k^2\langle\partial_k v_k|\partial_k v_k\rangle
\right].
\eeq
Inserting $1=P_k+Q_k$ in the last term on the RHS and then combining 
with Eq.~(\ref{eq:x_avg}) yields
\beq
\label{spread}
(\Delta x)^2=\frac{2\pi}{a}\int dk (\partial_k f_k)^2 
            +\langle G_k\rangle
            +\left<(\Delta A_k)^2\right>,
\eeq
where $G_k$ is the quantum metric,\cite{MV}
\beq
\label{eq:metric}
G_k=\langle\widetilde{\partial}_k v_k|\widetilde{\partial}_k v_k\rangle.
\eeq

All three terms in Eq.~(\ref{spread}) are manifestly non-negative. The first one
only depends on $f_k$, while the remaining two also depend
on the states $\ket{v_k}$. However, the second term 
is insensitive to the phases of those states (it is invariant
under the gauge transformation (\ref{eq:phase_trans})), whereas the third
term is phase-dependent.

It is instructive to consider
the limit of a uniform waveform, $f_k=a/2\pi$, 
in which $\ket{\phi}$ becomes a Wannier function.
Eq.~(\ref{eq:x_avg}) can then be recast as 
$\langle x\rangle=a\varphi/2\pi$, where
$\varphi=\int dk\,A_k$ is the Berry phase associated with the
manifold of states $\ket{v_k}$.\cite{KSV}
The first term on the RHS of Eq.~(\ref{spread}) then
vanishes identically, while the second and third terms reduce
to the gauge-invariant and gauge-dependent parts of the Wannier spread 
for an isolated band in one dimension, given respectively by
Eqs.~(C12) and (C17) of Ref.~\onlinecite{MV}.

\subsection{Uncertainty relation and minimal wavepackets}
\label{sec:uncertainty}

An alternative decomposition of the wavepacket spread may be 
obtained by noting that 
\beq
\big<\left(iD_k\right)^2\big>=
\frac{2\pi}{a}\int dk (\partial_k f_k)^2+
\langle A_k^2\rangle,
\eeq
as can be readily verified using the hermiticity of
$iD_k$ and the reality of $f_k$. Comparison with 
Eqs.~(\ref{eq:x_avg}) and (\ref{spread}) shows that
\beq
\label{Deltax}
(\Delta x)^2=
\left<(\Delta (iD_k))^2\right>
+\langle G_k\rangle.
\eeq
Combining this with the relation
\beq
(\Delta A)^2(\Delta B)^2\geq
\frac{1}{4}\left|\langle [A,B] \rangle\right|^2
\eeq
yields,  upon setting $A=iD_k$, $B=k$, and using $[iD_k,k]=i$, 
\beq
\label{eq:uncertainty}
\left[
(\Delta x)^2-\langle G_k\rangle
\right]
(\Delta k)^2\geq\frac{1}{4}.
\eeq
In the limit of a vanishing lattice potential $G_k\rightarrow 0$ and
$k\rightarrow p$ (canonical momentum). Eq.~(\ref{eq:uncertainty}) then reduces
the familiar Heisenberg uncertainty relation.

Let us now show that (\ref{eq:uncertainty}) becomes an
equality for minimal wavepackets in one dimension. 
Once the manifold of states $\ket{v_k}$ and the 
width $\Delta k$ of the waveform are specified, all that remains is to 
set the phases of the $\ket{v_k}$ and the shape of $f_k$. We wish to minimize 
the spread (\ref{spread}) with respect to those two parameters. We start with 
the phases, which only affect the term $\left<(\Delta A_k)^2\right>$. 
This term vanishes when $A_k$ is constant,\cite{explan-ak} in which case
\beq
(\Delta x)^2-\langle G_k\rangle=
\frac{2\pi}{a}\int dk (\partial_k f_k)^2.
\eeq
Combining the previous two equations,
\beq
\label{eq:waveform_inequality}
\frac{2\pi}{a}\int dk (\partial_k f_k)^2\geq\frac{1}{4(\Delta k)^2}.
\eeq
It can be verified that for $\Delta k\ll 2\pi/a$
this becomes an equality when the waveform has a Gaussian shape.
In conclusion, a minimal wavepacket in one dimension 
is characterized by
a Gaussian-shaped $f_k$ and a constant Berry connection $A_k$ in the
region of $k$-space where $f_k$ is non-negligible. Its spread equals
\beq
\label{eq:maxloc}
(\Delta x)^2_{\rm min}=\frac{1}{4(\Delta k)^2}+\langle G_k\rangle.
\eeq
It was shown in Ref.~\onlinecite{MV} that the spread of a maximally-localized
Wannier function in one dimension is $(\Delta x)^2_{\rm min}=\langle G_k\rangle$.
This result can be viewed as the limit $\Delta k\rightarrow\infty$ of 
Eq.~(\ref{eq:maxloc}).
In the opposite limit of a narrow waveform, the wavepacket spread 
becomes dominated by the term
$1/4(\Delta k)^2$, as will be illustrated in the next Section.

\section{Numerical Results}
\label{sec:results}

\subsection{Tight-binding model}

\begin{figure}[h]
\includegraphics[width=0.45\textwidth]{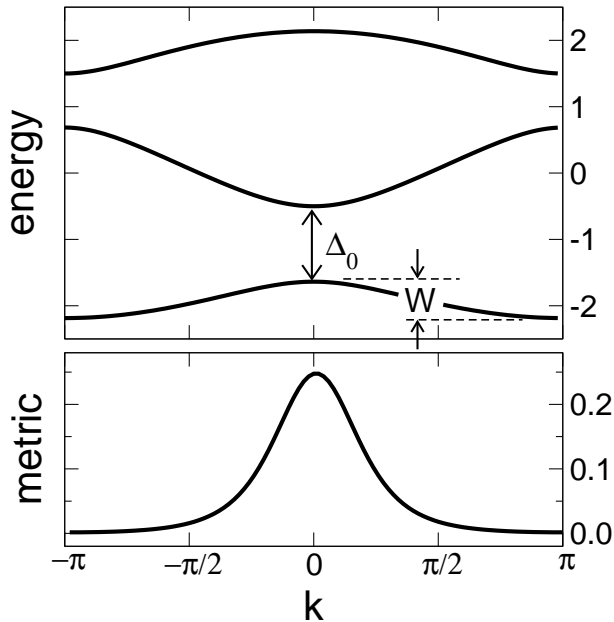}
\caption{
Upper panel: Band structure of the
one-dimensisonal tight-binding model of Eq.~(\ref{hamiltonian}), for the
choice of parameters $\gamma=-U=1$.
Lower panel: Quantum metric [Eq.~(\ref{eq:metric})] for the
lowest band.}
\label{bands} 
\end{figure}

We have applied our scheme to the same one-dimensional tight-binding model
used in Ref.~\onlinecite{SIV}. This is a three-band Hamiltonian with three
atoms per unit cell of length $a=1$ and one orbital per atom,
\beq
\label{hamiltonian}
H^0=\sum_j\Big\{
\epsilon_j c_j^\dagger c_j+
\gamma\big[c_j^\dagger c_{j+1}+c_{j+1}^\dagger c_j \big]
\Big\},
\eeq
with the site energy given by $\epsilon_{3m+l}=U\cos\beta_l$. Here $m$ is
the cell index, $l=\{-1,0,1\}$ is the site index, and $\beta_l=2\pi l/3$.
The upper panel of Fig.~\ref{bands} shows the energy dispersion for 
$\gamma=-U=1$.

Before the spatial distribution of the wavepacket can be defined, 
the matrix elements of the position operator must be 
specified. As in Ref.~\onlinecite{SIV} we choose the simplest diagonal 
representation $x=\sum_j x_j c_j^\dagger c_j$, with $x_j=j/3$. 
The lower panel of Fig.~\ref{bands} shows the 
quantum metric calculated for the Bloch states in 
the lowest band. As expected from the relation\cite{souza00}
$G_k\leq1/2\Delta_k$
($\Delta_k$ is the direct gap to the second band), $G_k$ peaks around $k=0$. 

We will now study numerically the wavepacket dynamics on this model.
In Secs.~\ref{sec:bo} and \ref{sec:zener}
we consider respectively Bloch oscillations
in a weak field and Zener
tunneling in a strong field. In both cases
we gradually turn on the field over a time interval, 
$T$, as $\ee(t)=\ee_0\sin(\pi t/2T)$, and keep it constant afterwards.
The simulations begin with a minimal wavepacket prepared in the lowest band.
Unless otherwise noted, the width of the Gaussian waveform is
$\Delta k=0.075\times 2\pi$.

\subsection{Bloch oscillations in a weak electric field}
\label{sec:bo}

\begin{figure}[h!]
\includegraphics[width=0.45\textwidth]{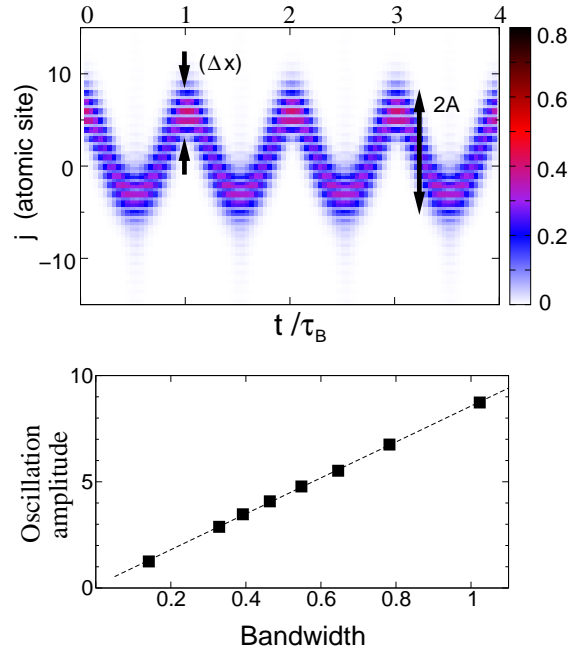}
\caption{(Color online.) Bloch oscillations of a wavepacket prepared
  in
the lowest band.
Upper panel: Time evolution of 
$|\langle j|\phi\rangle|^2$, the weights of the wavepacket
on the tight-binding basis orbitals, for tight-binding parameters 
$\gamma=-U=1$.
Lower panel: Amplitude $A$ of the Bloch oscillations versus the 
bandwidth $W$.
}
\label{BOs}
\end{figure}

In order to observe long-lived Bloch oscillations we choose a weak field 
$\ee_0=0.055\ll\Delta_0/a$, which we turn on over a time interval
of the order of the Bloch oscillation period
$\tau_B=2\pi/\ee_0a$ (henceforth in this subsection we choose $t=0$ 
long after the field has saturated at $\ee_0$). 
100 $k$-points are used to sample the Brillouin zone, and the time step is
$\Delta t=1.7\times 10^{-5}\tau_B$. 

In the upper panel of Fig.~\ref{BOs} the weights 
$|\langle j|\phi\rangle|^2$  of the wavepacket on the tight-binding 
orbitals $\ket{j}$ are
used to depict its spatial distribution as a function of time.  The 
Bloch oscillations of $\langle x\rangle$ are clearly seen. 
In the lower panel we plot the 
oscillation amplitude, $A$, versus the width, $W$, of the lowest 
band, which was tuned by adjusting the tight-binding parameters in the range 
$0.5\leq\gamma\leq1.5$ and $-1.5\leq U\leq -0.5$.
$A$ and $W$ are linearly related, with a slope
$A/W\simeq 8.5$. This is in agreement with the 
prediction $A=(1/2)(W/\ee_0)|S_0|$, valid for a single-band 
tight-binding model.\cite{BouchardLuban}
The dimensionless parameter $S_0$ depends on the initial 
choice of wavepacket,
but its magnitude cannot exceed $1$; in the present case
$1/2\ee_0\simeq 9.09$, so that $|S_0|\simeq 0.9$.

\begin{figure}[h]
\includegraphics[width=0.45\textwidth]{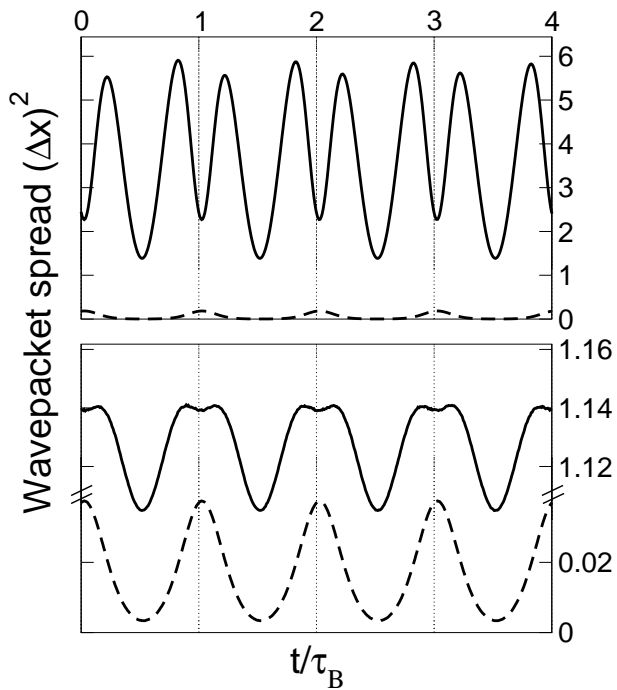}
\caption{Time evolution of the total
wavepacket spread $(\Delta x)^2$ (solid lines), and of the metric 
contribution $\langle G_k\rangle$ (dashed lines), for 
two different sets of tight-binding parameters.  
Upper panel: $\gamma=-U=1$, resulting in a bandwidth $W\sim 0.5$ and 
a minimum bandgap $\Delta_0\sim 1.1$. 
Lower panel: $\gamma=0.5$ and $U=-1$, for which 
$W\sim 0.1$ and $\Delta_0\sim 1.2$. 
}
\label{oscillations}
\end{figure}
Next we analyze the behavior of the 
wavepacket spread in the course of the Bloch oscillations, 
by tracking each of the three terms in Eq.~(\ref{spread}).
According to Eq.~(\ref{formevolution}), as
the center $\langle k\rangle$ of the packet traverses the 
Brillouin zone, the shape of the waveform $f_k$ remains unchanged. Hence the 
term $(2\pi/a)\int dk (\partial_k f_k)^2$
is a constant of motion. Oscillations in the spread must therefore arise from
the other two terms, $\langle G\rangle$ and $\left<(\Delta A)^2\right>$
(henceforth the $k$ subscript will be ommited for brevity).

\begin{figure}[h]
\includegraphics[width=0.45\textwidth]{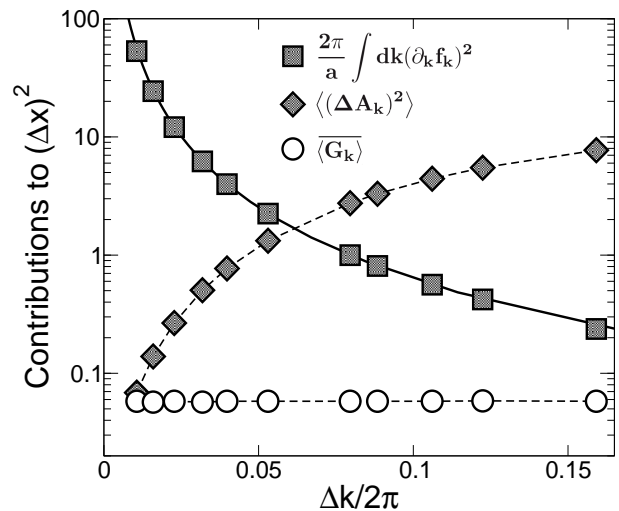}
\caption{Dependence of the time-averaged contributions to the wavepacket spread
[Eq.~(\ref{spread})] on the waveform width $\Delta k$, for $\gamma=-U=1$. 
The two dashed lines are fits to the values from the numerical
simulation (symbols), while the solid line equals $1/4(\Delta k)^2$,
the right-hand-side of Eq.~(\ref{eq:waveform_inequality}).}
\label{fig:spread}
\end{figure}

Since initially the wavepacket was minimal ($\left<(\Delta A)^2\right>=0$),
one might have expected spread oscillations to arise mostly
from the $k$-space dispersion of the metric, i.e., from
the varying constraint imposed
by the uncertainty relation (\ref{eq:uncertainty}) as the wavepacket
moves through $k$-space. Instead we find that 
after an initial transient the spread of the packet can become far from minimal,
with $\overline{\left<(\Delta A)^2\right>} \gg 
\overline{\langle G\rangle}$ (the bars denote a time average over several
Bloch oscillations). This is the situation depicted in the upper panel of
Fig.~\ref{oscillations}, which pertains to the
same simulation run as in the upper panel of 
Fig.~\ref{BOs}. Note that the spread undergoes sizeable oscillations,
arising mostly from $\left<(\Delta A)^2\right>$ (not shown).

The above scenario is typical of Bloch oscillations in a
relatively wide band. When we adjust the tight-binding parameters so as
to reduce the bandwidth, we find that
the wavepacket remains close to minimal in the course of the Bloch 
oscillations, and the associated spread oscillations are very weak,
of the order of the Brillouin zone
dispersion of the metric (lower panel of Fig.~\ref{oscillations}).
One could try to enhance the spread oscillations by further tuning the model
parameters so as to make the metric
very large in some regions of $k$-space. However, that would require
very small gaps, and under those circumstances the Bloch oscillations are
strongly damped by Zener tunneling.

So far we have considered a fixed waveform width 
$\Delta k=0.075\times 2\pi$.
Fig.~\ref{fig:spread} displays the dependence on $\Delta k$ of each
contributions to $(\Delta x)^2$, averaged over several Bloch oscillations. 
The term $\overline{\langle G\rangle}$ is roughly constant and equal to
the Brillouin zone average of the metric;
it remains a small fraction of  $(\Delta x)^2$ over the entire range of 
$\Delta k$. For $\Delta k\ll 2\pi/a$ the spread is dominated by
the term $(2\pi/a)\int dk (\partial_k f_k)^2$,
while for larger values of $\Delta k$ the term
$\overline{\left<(\Delta A)^2\right>}$ takes over. Its monotonic
increase is easily understood: the larger the range 
$\Delta k$, the
larger the spread of $A_k$ over that range is likely to be.

\subsection{Zener tunneling in a strong electric field}
\label{sec:zener}

\begin{figure}[h!]
\includegraphics[width=0.45\textwidth]{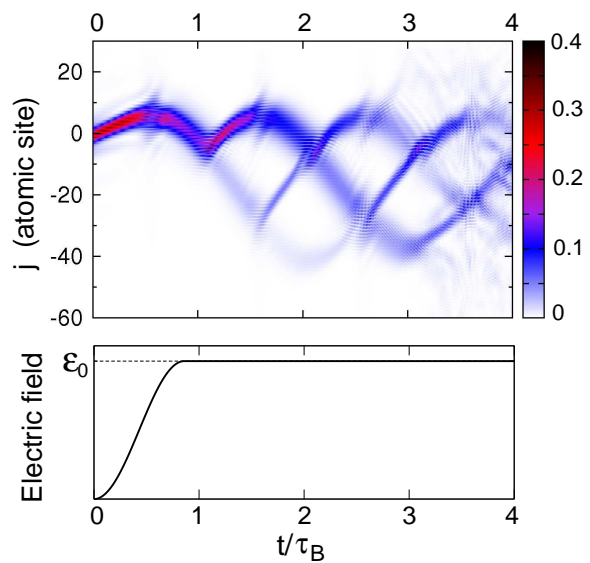}
\caption{(Color online.) 
Upper panel: Time evolution of a wavepacket prepared in the
lowest band,
with child packets in the second and third bands 
appearing as a result of Zener tunneling.  
The tight-binding parameters are $\gamma=1$ and $U=-0.3$.
Lower panel: Electric field as a function of time.
The field is switched on from $t=0$ to $t=0.86\tau_B$, where
$\tau_B$ is the Bloch period for the saturation field $\ee_0$.
}
\label{BO-zener}
\end{figure}
In the previous subsection a weak electric field 
($\ee_0\ll\Delta_0/a$) was chosen, so that for 
the duration of the simulation the wavepacket remained mostly in the
lowest band. For sufficiently strong fields, significant interband 
transitions are expected to occur as the wavepacket reaches the zone
center, where the gap is smallest.  
In order to observe this phenomenom  the saturation field was increased 
from $\ee_0=0.055$ to $\ee_0=0.09$, while the minimum gap between the 
first two bands was reduced from $\Delta_0=1.14$ to  $\Delta_0=0.31$
by setting $U=-0.3$.
With this choice of parameters a very good stability of the propagation
algorithm is needed, and we decreased the time step from $2\times 10^{-3}$ to 
$2\times 10^{-4}$. The Brillouin zone was sampled over 300 $k$-points.

The Zener tunneling can be seen in the upper panel of Fig.~\ref{BO-zener}
as a splitting of the wavepacket in real space,
At the end of every Bloch oscillation, $t=n\cdot\tau_B$ ($n=1,2,\dots)$, the
main wavepacket in the lowest band spawns child packets 
which oscillate in the second band with the
same period $\tau_B$ but a larger amplitude (due to the larger width of the
second band) and gain more weight after each Bloch cycle.
These child packets also spawn grandchild packets at 
$t=(2n+1)\tau_B/2$, when Zener tunneling from the second to the 
third band becomes possible.

\begin{figure}[t!]
\includegraphics[width=0.45\textwidth]{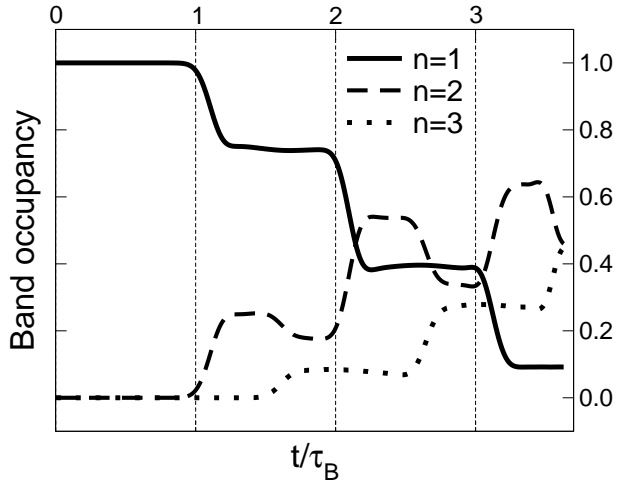}
\caption{Band occupancy $P_n(t)$ for the simulation
shown in Fig.~\ref{BO-zener}. 
}
\label{probability}
\end{figure}

A more quantitative picture is obtained by monitoring the 
distribution of the packet among the three bands.
We define the band occupancy $P_n(t)$ as the total 
probability that the wavepacket resides on band $n$:\cite{BouchardLuban}
\beq
P_n(t) = 
\sum_k\, f_k^2(t)\big|\big< u_{nk}^{(0)}\big|v_k(t)\big>\big|^2.
\eeq
This quantity is plotted in Fig.~\ref{probability} as a function of time.
Initially only the first band is occupied.  
After a Bloch
period, the wavepacket reaches the zone center, 
where the gap to the
second band is smallest, at which point significant Zener tunneling occurs, 
giving rise to a partial occupation of the second band.
(Between $t=0$ and $t=\tau_B$ the wavepacket moved in $k$-space 
by less than the full Brillouin zone width, from
$\langle k\rangle=-2\pi/3$ to $\langle k\rangle\simeq -2\pi$, because during 
most of that time interval
the electric field strength was less than $\ee_0$, as shown in
the lower panel of Fig.~\ref{BO-zener}.)
Subsequently there are also transitions to the 
third band, again with periodicity $\tau_B$;
because transitions from the second band to the first and third bands happen
at the zone center and at the zone boundary respectively, $P_2$ 
undergoes changes
twice as often as $P_1$ and $P_3$.

\section{Conclusions}

In this work we have developed a new numerical scheme for simulating
wavepacket dynamics in a periodic lattice potential with a linear potential
(homogeneous electric field) superimposed. By using the $k$-space 
representation of the position operator, we were able to include the linear
electric-field term $e\boldsymbol{\ee}\cdot{\bf r}$ in the Hamiltonian
under periodic boundary conditions, thus avoiding having to use large
supercells with hard-wall boundary conditions.\cite{BouchardLuban}

In the present approach
the wavepacket is represented on a uniform mesh of $k$-points by a waveform 
$f_k$ sitting on top of a ``band'' of states 
$\ket{v_k}$ [Eq.~(\ref{wavepacket})]. The time evolution of the wavepacket is
then obtained from that of $f_k$ and $\ket{v_k}$. For $\ee\not=0$ the states
$\ket{v_k}$ 
become non-adiabatic, field-polarized Bloch states which span
several energy bands of the field-free crystal Hamiltonian $H^0$ 
(a similar representation of wavepackets
in coupled energy bands was used in Ref.~\onlinecite{culcer05}); thus
interband effects such as Zener tunneling are fully accounted for.

The method was tested on a one-dimensional tight-binding model. 
Depending on the choice of tight-binding parameters and electric field 
strength, we observed either long-lived Bloch oscillations, or short-lived 
Bloch oscillations strongly damped by Zener tunning. In the former regime
we monitored the changes in the wavepacket spread accompanying the Bloch
oscillatations of the center of mass, and identified two distinct situations:
(i) For wavepackets moving in narrow bands, the spread changed very 
little over time.
(ii) For wavepackets moving in wide bands, the Bloch oscillations were 
accompanied by considerable oscillations of the wavepacket spread. 
An analysis of the $k$-space expression for the spread 
[Eq.~(\ref{spread})]
reveals
two distinct contributions which can change over time: one associated with
the Berry connection of the underlying Bloch states, and another related to the
quantum metric. By tracking each of them separately, we 
concluded that in the cases where significant spread oscillations took place, 
they originated mostly in the Berry connection term.

\begin{acknowledgments}
We thank David Vanderbilt for useful discussions. This work was supported by
the Marie Curie OIF program of the EU and by NSF Grant No.~DMR-0706493.
\end{acknowledgments}

\appendix*

\section{
Discretized expressions in $k$-space
}
\label{app:finite_diff}

Here we derive the discretized versions used in Sec.~\ref{sec:results}
of the dynamical equations of Sec.~\ref{sec:method} and of the 
wavepacket center and spread expressions of
Sec.~\ref{sec:center_spread}.

\subsection{Dynamical equations}

The appropriate finite-difference representation of
$\ket{\widetilde{\partial}_k \vv_k}$ on a uniform grid is\cite{SIV}
\beq
\label{eq:disc_cov}
\ket{\widetilde{\partial}_k \vv_k}=\frac{1}{2b}
\left(\ket{\widetilde{v}_{k+b}}-\ket{\widetilde{v}_{k-b}}\right),
\eeq
where $b$ is the mesh spacing and 
\beq
\ket{\widetilde{v}_{k+b}}=
\frac{\ket{\vv_{k+b}}}{\bra{\vv_k}\vv_{k+b}\rangle}.
\eeq
We use Eq.~(\ref{eq:disc_cov}) to recast 
Eq.~(\ref{state2evolution}) as\cite{SIV}
\beq
\label{eq:tdse_b}
i\ket{\dot{\vv}_k}=T_k\ket{\vv_k},
\eeq
in terms of the Hermitian operator
\beq
T_k=\h0+w_k+w_k^\dagger,
\eeq
where $w_k=(i\ee/2b)\left(P_k^+-P_k^-\right)$ and
$P_k^\pm=\ket{\widetilde{v}_{k\pm b}}\bra{\vv_k}$.
Eq.~(\ref{eq:tdse_b}) is solved numerically at each grid point 
using\cite{BouchardLuban,SIV}
\beq
\label{eq:time_propag}
\ket{\vv_k}(t+\Delta t)\simeq
\frac{1-i(\Delta t/2)T_k(t)}{1+i(\Delta t/2)T_k(t)}\ket{\vv_k}(t).
\eeq

Because of the similarity between Eqs.~(\ref{phaseevolution})
and (\ref{eq:tdse_b}), the phase factors $U_k$ can be propagated in time using 
the same algorithm. A finite-difference representation for the connection 
in Eq.~(\ref{phaseevolution}) is needed. We use\cite{MV} 
\beq
\label{eq:A_disc}
A_k\simeq -\frac{1}{2b}
\left(
\Phi_k^+-\Phi_k^-
\right),
\eeq
where we have defined the phases
$\Phi_k^{\pm}=-{\rm Im}\ln\langle v_k|v_{k\pm b}\rangle$.
When propagating $U_k$ via Eq.~(\ref{phaseevolution}),
care must be taken in choosing consistently the branch cuts for the two 
phases in Eq.~(\ref{eq:A_disc}), to ensure that $A_k$ remains a smooth 
function of $k$ at every time step.

\subsection{Wavepacket center and spread}

In order to obtain a finite-difference representation of Eq.~(\ref{eq:x_avg}) 
we make the replacement $\int dk\rightarrow\sum_k\,b$
and then use Eq.~(\ref{eq:A_disc}). A few manipulations yield
\beq
\label{eq:x_avg_disc}
\langle x\rangle\simeq -\frac{2\pi}{a}\sum_k\,
\frac{1}{2}\left(f_k^2+f_{k+b}^2\right)\Phi_k^+.
\eeq
To find an expression for $(\Delta x)^2$ 
we start from Eq.~(\ref{spread}). The first 
term on the RHS is easily discretized, and $\langle G_k\rangle$
can be evaluated from (\ref{eq:metric}).
The remaining term is $\langle A_k^2\rangle-\langle A_k\rangle^2$.
For $\langle A_k\rangle^2=\langle x\rangle^2$ we 
use (\ref{eq:x_avg_disc}) and, from (\ref{eq:A_disc}),
\bea
\label{eq:omega_d_disc}
\langle A_k^2\rangle\simeq
\frac{\pi}{b}\sum_k\,
\left[
\frac{f_k^2+f_{k+b}^2}{2}(\Phi_k^+)^2-
f_k^2\Phi_k^+\Phi_k^-
\right].
\eea

Besides reducing to the correct continuum expressions as
$b\rightarrow 0$, Eqs.~(\ref{eq:x_avg_disc}) 
and (\ref{eq:omega_d_disc}) preserve exactly, for finite $b$, certain 
properties of those expressions. If we perform a change of phases
\beq
\label{eq:phase_trans}
\ket{v_k}\rightarrow e^{i\theta_k}\ket{v_k}
\eeq
with $\theta_k=\gamma_k-kR$ ($R$ is a lattice vector and
$\gamma_{k+2\pi/a}=\gamma_k$),
the center of the packet, Eq.~(\ref{eq:x_avg}), changes as
\beq
\label{eq:xgauge_b}
\langle x\rangle\rightarrow\langle x\rangle+R-
\frac{2\pi}{a}\int dk\, f_k^2 \partial_k\gamma_k.
\eeq
For a Wannier wavepacket $f_k$ is constant, so that the last term vanishes, and
$\langle x\rangle$ changes at most  by a lattice vector.\cite{KSV}
If instead $f_k$ spans a 
narrow region ${\cal K}$ of the Brillouin zone, $\langle x\rangle$
can then shift continuously under (\ref{eq:phase_trans}). 
Consider the choice
\beq
\label{eq:shift}
\theta_k=
\begin{cases}
-\delta k & k\in {\cal K}\\
0 & \mbox{otherwise}
\end{cases},
\eeq
which produces a rigid shift 
$\langle x\rangle\rightarrow\langle x\rangle+\delta$.
Eq.~ (\ref{eq:x_avg_disc}) 
obeys this transformation exactly, while the spread 
evaluated using (\ref{eq:omega_d_disc}) remains unchanged,
as can be easily verified.

\bibliography{pap.bib}

\begin{thebibliography}{18}
\bibitem[{exp()}]{explan-ak}
\bibinfo{note}{In practice a constant $A_k$ can be achieved on a uniform mesh
  following the prescription in Sec.~III.C.1 of Ref.~\onlinecite{MV}.}
\end{thebibliography}

\end{document}